\documentclass[aps,prl,preprint,tightenlines,superscriptaddress,showpacs,byrevtex]{revtex4}

\usepackage{graphicx}% Include figure files
\usepackage{dcolumn}% Align table columns on decimal point
\usepackage{bm}% bold math
\usepackage{xspace} 
\usepackage{amsmath}
\usepackage{amssymb}

\newcommand{\dzerdm}{\ensuremath{B^- \to D^0 D^-}\xspace}
\newcommand{\dzdstm}{\ensuremath{B^- \to D^0 D^{*-}}\xspace}
\newcommand{\dpludm}{\ensuremath{B^0 \to D^+ D^-}\xspace}
\newcommand{\dzedsm}{\ensuremath{B^- \to D^0 D_s^-}\xspace}

\def\to {\rightarrow}

\newcommand{\CP}{\emph{CP}\xspace}
\newcommand{\de}{\ensuremath{\Delta E}\xspace}
\newcommand{\mb}{\ensuremath{M_\mathrm{bc}}\xspace}

\begin{document}

%\title{
% Observation of $B^0 \to D^+ D^-$, $B^- \to D^0 D^-$ and $B^- \to D^0 D^{*-}$ decays
% \Large \bf Observation of $\dpludm$, $\dzerdm$ and $\dzdstm$ decays}

\title{\Large \bf Observation of \boldmath{\dpludm},
  \boldmath{\dzerdm} and \boldmath{\dzdstm} decays}

\begin{abstract}
        We report the first observation of the decay modes \dpludm,
        \dzerdm and \dzdstm based on 152 $\times$ 10$^6$
        $B\overline{B}$ events collected at KEKB.
        The branching fractions of 
        $\dpludm$, $\dzerdm$ and $\dzdstm$  are found to be
        $(3.21 \pm 0.57 \pm 0.48) \times 10^{-4}$,
        $(5.62 \pm 0.82 \pm 0.65) \times 10^{-4}$ and  
        $(4.59 \pm 0.72 \pm 0.56) \times 10^{-4}$, respectively.
        Charge asymmetries in the $\dzerdm$ and $\dzdstm$ 
        channels are consistent with zero.
       
\end{abstract}

\affiliation{Budker Institute of Nuclear Physics, Novosibirsk}
\affiliation{Chiba University, Chiba}
\affiliation{Chonnam National University, Kwangju}
\affiliation{University of Cincinnati, Cincinnati, Ohio 45221}
\affiliation{University of Hawaii, Honolulu, Hawaii 96822}
\affiliation{High Energy Accelerator Research Organization (KEK), Tsukuba}
\affiliation{Hiroshima Institute of Technology, Hiroshima}
\affiliation{Institute of High Energy Physics, Chinese Academy of Sciences, Beijing}
\affiliation{Institute of High Energy Physics, Vienna}
\affiliation{Institute for Theoretical and Experimental Physics, Moscow}
\affiliation{J. Stefan Institute, Ljubljana}
\affiliation{Kanagawa University, Yokohama}
\affiliation{Korea University, Seoul}
\affiliation{Kyungpook National University, Taegu}
\affiliation{Swiss Federal Institute of Technology of Lausanne, EPFL, Lausanne}
\affiliation{University of Ljubljana, Ljubljana}
\affiliation{University of Maribor, Maribor}
\affiliation{University of Melbourne, Victoria}
\affiliation{Nagoya University, Nagoya}
\affiliation{Nara Women's University, Nara}
\affiliation{National Central University, Chung-li}
\affiliation{National United University, Miao Li}
\affiliation{Department of Physics, National Taiwan University, Taipei}
\affiliation{H. Niewodniczanski Institute of Nuclear Physics, Krakow}
\affiliation{Nihon Dental College, Niigata}
\affiliation{Niigata University, Niigata}
\affiliation{Osaka City University, Osaka}
\affiliation{Osaka University, Osaka}
\affiliation{Panjab University, Chandigarh}
\affiliation{Peking University, Beijing}
\affiliation{Saga University, Saga}
\affiliation{University of Science and Technology of China, Hefei}
\affiliation{Seoul National University, Seoul}
\affiliation{Sungkyunkwan University, Suwon}
\affiliation{University of Sydney, Sydney NSW}
\affiliation{Tata Institute of Fundamental Research, Bombay}
\affiliation{Toho University, Funabashi}
\affiliation{Tohoku Gakuin University, Tagajo}
\affiliation{Tohoku University, Sendai}
\affiliation{Department of Physics, University of Tokyo, Tokyo}
\affiliation{Tokyo Institute of Technology, Tokyo}
\affiliation{Tokyo Metropolitan University, Tokyo}
\affiliation{Tokyo University of Agriculture and Technology, Tokyo}
\affiliation{University of Tsukuba, Tsukuba}
\affiliation{Virginia Polytechnic Institute and State University, Blacksburg, Virginia 24061}
\affiliation{Yonsei University, Seoul}
   \author{G.~Majumder}\affiliation{Tata Institute of Fundamental Research, Bombay} % Tata
   \author{K.~Abe}\affiliation{High Energy Accelerator Research Organization (KEK), Tsukuba} % KEK
   \author{K.~Abe}\affiliation{Tohoku Gakuin University, Tagajo} % TohokuGakuin
   \author{I.~Adachi}\affiliation{High Energy Accelerator Research Organization (KEK), Tsukuba} % KEK
   \author{H.~Aihara}\affiliation{Department of Physics, University of Tokyo, Tokyo} % Tokyo
   \author{Y.~Asano}\affiliation{University of Tsukuba, Tsukuba} % Tsukuba
   \author{V.~Aulchenko}\affiliation{Budker Institute of Nuclear Physics, Novosibirsk} % BINP
   \author{T.~Aushev}\affiliation{Institute for Theoretical and Experimental Physics, Moscow} % ITEP
   \author{S.~Bahinipati}\affiliation{University of Cincinnati, Cincinnati, Ohio 45221} % Cincinnati
   \author{A.~M.~Bakich}\affiliation{University of Sydney, Sydney NSW} % Sydney
% ...
% Ljubljana
\collaboration{The Belle Collaboration}

\maketitle

Mixing induced \CP-violating asymmetries in $b \to c \overline{c} s$ decays
have been observed at the B factory experiments, Belle and BaBar,
at levels consistent with Standard Model (SM) predictions \cite{cpsm}.
Cabibbo-suppressed double charm decays (e.g. $B^0 \to D^{(*)+}D^{(*)-}$)
are dominated by $b \to c \overline{c} d$ tree diagram contributions. Additional
penguin contributions with a different weak phase are expected to
be small in comparison with tree diagram contributions. Hence, time 
dependent \CP-violating asymmetries in such
double charm decays should be nearly equal to those in 
$B^0 \to J/\psi K_S$ ($b \to c \overline{c} s$) type decays. However, a 
variety of processes beyond
the SM can provide additional sources of \CP violation \cite{beyond}. Thus, 
the $B^0 \to D^{(*)+}\,D^{(*)-}$ decay modes can
be used to confirm the SM predictions of \CP violation \cite{sanda} or to 
look for physics beyond the Standard Model.

Signals for the decay modes $B^0 \to D^{*+}D^{*-}$, $B^0 \to D^{*+}D^{-}$ 
and $B^0 \to D^{+}D^{*-}$ have already been observed. %\cite{dstdst} 
Measurements of time dependent \CP-violating 
asymmetry parameters are also available for these modes \cite{cpvxx}.
While the branching fraction for \dpludm is expected to be fairly
large, evidence for this decay mode has not yet been reported.
Using SU(3) symmetry and the world-average branching fraction \cite{pdg2004}, 
${\cal{B}}(B^0 \to D^{+} D^{-})$ is estimated to be
$\sin^2\theta_{\rm c} \times{\cal{B}}(B^0 \to D_s^{+} D^{-})$ 
$\simeq (4.0 \pm 1.5) \times 10^{-4}$, where 
$\theta_{\rm c}$ is the Cabibbo angle.

The decay modes $\dzerdm$ and  $\dzdstm$ are expected to be dominated by 
tree diagrams with some additional contributions from penguin diagrams.
Scaling from the well-measured branching fractions for the Cabibbo-favored
processes $B^- \to D^0 D_s^-$ and 
$B^- \to D^0 D_s^{*-}/B^- \to D^{*0} D_s^{-}$, one can estimate the
branching fractions of these two decay modes to be 
$(6.5 \pm 2.0) \times 10^{-4}$ and $(5.1 \pm 2.2) \times 10^{-4}$, 
respectively. Assuming SU(3) symmetry, measurement of these branching 
fractions will enable better understanding of the penguin processes.

Here, we report the first observation of the decays $\dpludm$,
$\dzerdm$ and $\dzdstm$. 
%Implicit charge conjugate modes are implied throughout this report.
Inclusion of charge conjugate modes is implied throughout this paper.
The analysis is based on a $140\,\mathrm{fb}^{-1}$ data sample
at the $\Upsilon(4S)$ resonance (10.58 GeV) 
and a 16 $\mathrm{fb}^{-1}$ data sample 60 MeV below the $\Upsilon(4S)$ peak 
(referred to as off-resonance data), collected with the Belle 
detector \cite{belledet} at the energy asymmetric $e^+e^-$ collider KEKB 
\cite{kekb}. The data 
sample contains  152 $\times 10^6$ $B\overline{B}$ events. 
The fractions of neutral and charged $B$ mesons produced in 
$\Upsilon(4S)$ peak are assumed to be equal.

The Belle detector is a general purpose magnetic spectrometer 
with a 1.5~T magnetic field provided by a superconducting solenoid.
Charged particles are measured using a 50 layer Central Drift Chamber 
(CDC) and a three layer double sided Silicon Vertex Detector (SVD). Photons 
are detected in an electromagnetic calorimeter (ECL) consisting of
8736 CsI(Tl) crystals. 
Exploiting the information acquired from an array of 128 time-of-flight
counters (TOF), an array of 1188 silica aerogel 
\v{C}erenkov threshold counters (ACC) and $dE/dx$-measurements in the CDC, we
derive particle identification (PID) likelihoods 
${\mathcal{L}}_{\pi/K}$.
A kaon candidate is identified by a requirement on the likelihood ratio
${\mathcal{L}}_K/({\mathcal{L}}_K+{\mathcal{L}}_\pi)$ such that the average
kaon identification efficiency is $\sim$ 93\% 
with pion misidentification rate of $\sim$ 10\%. 
Similarly, charged pions are selected with an efficiency of $\sim$ 95\% 
and kaon misidentification rate of $\sim$ 10\%.
We select charged pions and kaons that originate from the region
$dr <$ 1\,cm and $|dz|<$ 4\,cm with 
respect to the run dependent interaction points (IP), where $dr$, 
$dz$ are the distances of closest approach of $\pi/K$ tracks 
to the IP in the plane perpendicular to
and along the $z$-axis (the $z$-axis is defined as passing through the
nominal interaction point and antiparallel to the positron beam).
All tracks compatible with the electron
hypothesis ($\sim$ 0.2\% misidentification rates from pion/kaon) are eliminated.
No attempt has been made to identify muons, which represent a
background of about 2.7\% to the pion tracks.

Neutral kaons ($K_S$) are reconstructed via the decay 
$K_S \to \pi^+ \pi^-$ with no
particle identification
requirement for daughter pions
and the two-pion invariant mass is required to be within 11 MeV/$c^2$ 
($\sim 3.5 \sigma$, where $\sigma$ is the invariant mass
resolution of $\pi^+ \pi^-$) of the $K_S$ mass. To improve the purity of
$K_S$ candidates, we impose $K_S$ momentum-dependent criteria
on the impact parameter of the pions, the distance between
the closest approaches of the pions along the beam direction, 
the distance of the $\pi^+\pi^-$ vertex from the 
interaction point, and the azimuthal angle difference between the 
direction of 
$\pi^+\pi^-$ vertex from the IP and the $K_S$ momentum direction.
Mass and vertex constrained fits are applied to obtain the 4-momenta of 
$K_S$ candidates.
Neutral pions ($\pi^0$) are reconstructed from pairs of isolated 
ECL clusters (photons) with invariant mass in
the window 118 MeV/$c^2$ $< M_{\gamma \gamma} <$ 150 MeV/$c^2$ 
($\sim \pm 3 \sigma$). The energy of each
photon is required to be greater than 30 MeV in the barrel region, defined
as $32^\circ < \theta_\gamma < 128^\circ$, and greater than 50 MeV in the
endcap regions, defined as $17^\circ < \theta_\gamma \leq 32^\circ$ or
$128^\circ < \theta_\gamma \leq 150^\circ$, where $\theta_\gamma$ denotes the 
polar angle of the photon. Mass constrained fits are applied to obtain the 
4-momenta of $\pi^0$ candidates.

Beam gas events are rejected using the requirements $|P_z| <$ 2\,GeV/$c$ 
and 0.5 $< E_{\rm vis}/\sqrt{s}<$ 1.25, in the $\Upsilon(4S)$ rest frame, 
where $P_z$ and $ E_{\rm vis}$ are the sum of the longitudinal momentum 
and the energy of all reconstructed particles, 
respectively, and $\sqrt{s}$ is the sum of the beam energies in the
$\Upsilon(4S)$ rest frame. The continuum  
($e^+e^- \to q\overline{q}$, where $q=u,d,s,c$) events are suppressed 
by requirements
on the ratio of the second to the zeroth Fox-Wolfram moments
\cite{foxwol}, $R_2 <$ 0.35.

The ${D}^0$ meson is reconstructed through its decay to 
$K^-\pi^+,\,K^-\pi^+\pi^0,\,K^-\pi^+\pi^+\pi^-, \,K_S \pi^+ \pi^-$ and 
$K^+ K^-$. The ${D}^+$ meson is reconstructed through its decay to
$K^- \pi^+ \pi^+$ and $K^-K^+\pi^+$.
Mass and vertex constrained fits 
are applied to improve the $D$ meson momentum resolution.
 
The tracks from the $D$ decays are chosen with a criterion, 
$|dz_{\rm i} - dz_{\rm j}| <$ 2 cm,  where $dz_{\rm i(j)}$ is the
distance of closest approach of track $i(j)$ to the IP along the $z$-axis.
Large combinatorial backgrounds
are removed by requiring that the invariant mass of daughter
particles is within 2.5 $\sigma$ from the nominal $D$-mass,
where $\sigma$ ($\sim$ 4.5 MeV/$c^2$), the mass resolution,
depends on the decay chain.

$D^{*+}$ candidates are reconstructed by combining
the ${D}^0$ with a slow charged pion with $dr <$ 2\,cm 
and $|dz|$ $<$ 10\,cm with respect to the $D$ vertex.
${D}^0$ mass windows are widened to $\pm$ 20 MeV$/c^2$ for the reconstruction of
$D^{*+}$ candidates.
$D^{*+}$ candidates are required to have a reconstructed mass
difference between the $D^{*+}$ and $D^0$ within
2.0~MeV$/c^2$ ($\sim 3.0 \sigma$) of the nominal mass difference.
A kinematic fit with 
the ${D}^{*+}$ mass is applied to obtain 
the 4-momenta of the ${D}^{*+}$ candidate.

To reduce large combinatorial backgrounds in ${D}^0$ decays to
$K^-\pi^+\pi^+\pi^-$, a tighter impact parameter requirement, $dr <$ 0.5 cm, 
is applied to all four tracks. Note that this is not applied for 
\mbox{$D^{*+} \to D^0 (\to K^-\pi^+\pi^+\pi^-) \pi^+$} signals 
because this mode has much less background.
The $D^0 \to K^-\pi^+\pi^0$ decay mode is used only for $D^{*+}$ 
candidates to avoid large backgrounds in $D^0$ signals in this
decay mode.

Combinations of $D \overline{D}{}^{(*)}$ are used to reconstruct
candidate $B$ mesons. $D \overline{D}{}^{(*)}$ signals are contaminated with 
background from misidentified $D$ mesons and also combination of two $D$ candidates 
from opposite $B$ mesons. There are two important kinematic variables to extract 
signals
from these backgrounds, 
(\emph{i}) the energy difference, \de, between the measured energy of the 
candidate event and the beam energy, $E_\mathrm{beam}$, in the $\Upsilon(4S)$ rest
frame and (\emph{ii}) the beam energy constrained mass, 
$\mb = \sqrt{E^2_\mathrm{beam} - (\sum_i \vec{P}_i)^2}$, where 
$ \vec{P}_{i}$ are momentum vectors of the primary $D^{(*)}$ candidates. 
The \de distribution is used to extract the signal yield since
peaking backgrounds are expected in the \mb distributions.
The fit is performed for events where \mb satisfies
5.272 GeV/$c^2$ $< \mb <$ 5.288 GeV/$c^2$ and the fit range in \de is from
$-$70 MeV to 200 MeV. The restricted range in negative \de is chosen to
exclude
contributions from other $B$ decays, such as 
$B^0 \to \overline{D{}^0}D^{*0} (D^{*0} \to D^0 \gamma)$.
Selected events contain multiple $B$ candidates with a multiplicity
depending on the signal channels, which varies from 1.02 to 1.07.
In events with more than one candidate $B$ meson, the candidate with the
smallest $\chi^2 (=
 (\Delta M_{{D_1}}/\sigma_{M_{D_1}})^2  + %\oplus
  (\Delta M_{{D_2}}/\sigma_{M_{D_2}})^2$) is chosen, where $\Delta M_{D_i}$ is 
the difference of the reconstructed and nominal mass of $D_i$, and 
$\sigma_{M_{D_i}}$ is the resolution in $M_{D_i}$. In \dzdstm 
candidates, a $(\Delta M_{D^{*-}}/\sigma_{M_{D^{*-}}})^2$ term is also
added to the $\chi^2$ to choose the best candidate, where 
$\Delta M_{D^{*-}}$ is the difference between measured
and nominal mass difference between $D^{*-}$ and $\overline{D}{}^0$ , 
and $\:\sigma_{M_{D^{*-}}}$ is the resolution in 
measured mass difference.

Unbinned extended maximum likelihood fits to \de distributions are used 
to extract the signal yields. The signal shape is modeled as
a sum of two Gaussians, % and background is modeled with a linear function,
\begin{equation*}
  \begin{split}
    F(\de)&=A \Biggl\lbrace
      \exp \left[ -0.5 \left( \frac{\de-\mu}{\sigma}\right)^2 \right] \\
        &\phantom{=} + f_1 
        \exp \left[ -0.5 \left( \frac{\de-\mu}{f_2 \sigma} \right)^2 \right]
      \Biggr\rbrace
  \end{split}
\end{equation*}
which is dominated by a Gaussian of width, $\sigma$ $\sim$ 6 
MeV and a wider Gaussian, whose width is $\sim$ 2--3 times larger than
the main Gaussian function, but whose contribution is only
15\%--17\% of the main Gaussian function. Backgrounds are modeled
with a linear function.

Signal Monte Carlo (MC) is fitted with this function to determine $f_1$ and $f_2$. 
The fitted values of $f_1$ and $f_2$ are 0.051 and 2.48, respectively. 
The \de distribution in data is wider than in MC. 
The decay channel, \dzedsm is used as a control sample to calculate the 
scaling factor,
$f~ (~=~[\sigma_{data}/\sigma_\mathrm{MC}]_{\dzedsm} = 1.08 \pm 0.05$ ) of the 
Gaussian width in data.

To extract signal yields from data,
the width of the main Gaussian function is fixed to $\sigma =
f \sigma_\mathrm{MC}$ and remaining four parameters (parameters 
$A$ and $\mu$ of signal function and two parameters of the linear background 
function) are determined in the fit.

Signal and backgrounds are studied with 
Monte Carlo event samples that are generated using the QQ event 
generator \cite{qq98}. The response
of the Belle detector is simulated by a GEANT3-based program
\cite{geant3}. The simulated events are reconstructed and analysed 
with the same procedure as is used for the real data. 
A large generic $B\overline{B}$ MC samples, with a luminosity equivalent 
to 330 $\mathrm{fb}^{-1}$ of data is used to look for peaking 
backgrounds in \de distributions.
We have also studied feed across among these signals and
from other $B \to \overline{D}{}^{(*)}\overline{D}{}^{(*)}$ decays.
Background due to continuum events are studied by analysing the 
16 $\mathrm{fb}^{-1}$ of off-resonance data and simulated MC events 
equivalent to $\sim$ 330 $\mathrm{fb}^{-1}$ of data. 
No peaking backgrounds are observed in \de distributions of 
these samples.
Signal efficiencies in $\dpludm$, $\dzerdm$ and $\dzdstm$ decay modes
are 11.9$\pm$0.1\%, 11.1$\pm$0.1\% and 4.8$\pm$0.1\%, respectively.

\begin{figure}[htbp]
\begin{center}
 \includegraphics[width=.45\textwidth]{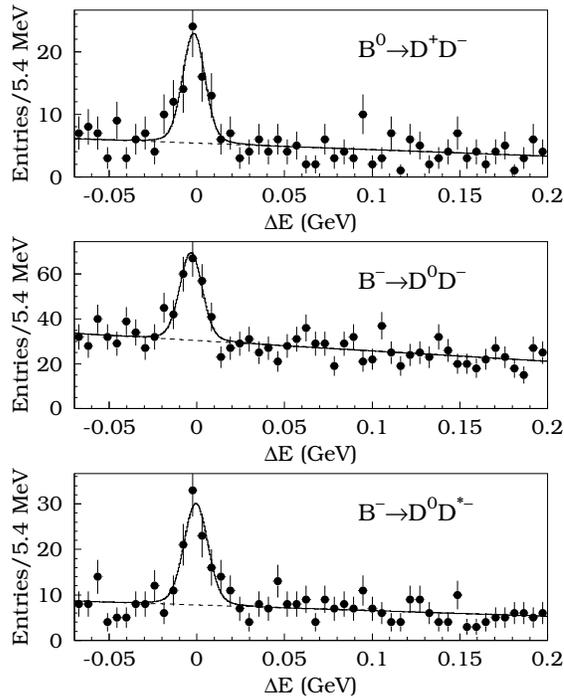}
\caption{Fit results of \de distributions in data.
         Points with error bars are the observed 
         events in data, solid lines are the results from the fit and 
         dashed lines represent the background components.}
 \label{fig:signal1}
\end{center}   
\end{figure}

\begin{table}[htbp] 
\begin{center}
\caption{Observed signal yields, statistical significances ($\sigma$)
 and branching fractions.}
\vspace{1mm}
\begin{ruledtabular}
\begin{tabular}{lrrrcrrrrr}
Channel & \multicolumn{3}{c}{$N_{\rm obs}$} & $\sigma$ &  \multicolumn{5}{c}{${\cal{B}} \times 10^4$}\\ \hline
$\dpludm $ & 54.3   &$\pm$&  9.7&  7.3&   3.21&$\pm$& 0.57&$\pm$&  0.48 \\
$\dzerdm $ & 120.5  &$\pm$& 17.6&  8.0&   5.62&$\pm$& 0.82&$\pm$&  0.65 \\ 
$\dzdstm $ & 73.6   &$\pm$& 11.5&  8.2&   4.59&$\pm$& 0.72&$\pm$&  0.56 \\
%$\dzedsm $ & 461.0 &$\pm$& 22.2& 36.9& 121.57&$\pm$& 5.85&$\pm$& 32.29 \\ \hline
%   \multicolumn{4}{|c|}{$\dzedsm$ : PDG}  &      \multicolumn{3}{c|}{130.0 $\pm$ 40.0} \\

\end{tabular} 
\end{ruledtabular}

\label{tab:branch}
\end{center}
\end{table}

Figure \ref{fig:signal1} shows the \de distributions in data
for the decay modes \dpludm, \dzerdm and \dzdstm.
There are clear structures near \de = 0. The results of the
fits for these decay modes are also shown in these plots. 
The signal yields 
obtained from the fits are given in Table \ref{tab:branch}.
The statistical significance of the yields, defined as 
$\sqrt{-2\,\ln\,({\cal{L}}_0/{\cal{L}}_{\max})}$, is 
7.3, 8.0 and 8.2 for the
$\dpludm$, $\dzerdm$ and $\dzdstm $ channels, respectively,
where ${\cal{L}}_0$(${\cal{L}}_{\max}$) is the maximum likelihood 
without (with) the signal contribution.
The corresponding \de distributions for events in the \mb 
sideband region (5.2 GeV/$c^2$ $< \mb <$ 5.26 GeV/$c^2$) are also
checked and they do not show any structure. 
To check for possible background from modes such as
$B\to D K\pi$, $B\to D K\pi\pi$
or charmless $B\to K \pi K (n\pi)$, we also examine the
$B$ signal yield for combinations when one of the $D$ candidates
has an invariant mass in a sideband outside the nominal $D$ mass window.
No significant yield is observed in such combinations.

Branching fractions obtained for these three modes are
listed in Table \ref{tab:branch}, where the first error is statistical
and the second error is systematic. This is the first measurement of 
the branching fractions for these decay modes. All results are 
consistent with the expectation from SU(3) symmetry.
As a consistency check, the \mb distributions are also fitted
after constraining $|\de| <$ 40 MeV and they show consistent 
signal yields.

\begin{figure}[htbp]
\begin{center}
 \includegraphics[width=.45\textwidth]{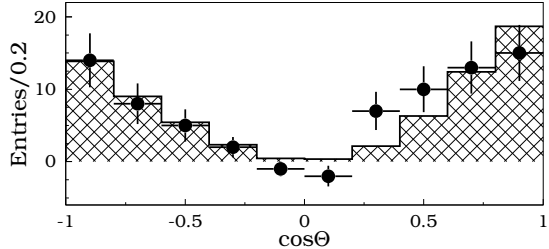}
\caption{Background-subtracted distribution of the cosine of the helicity 
         angle in $D^{*-} \to \overline{D}{}^0 \pi^-$ decay for \dzdstm
         candidates in data (points with error bars) and for 
         \dzdstm signal MC (hatched histogram).}
 \label{fig:helang}
\end{center}   
\end{figure}

The distribution of the helicity angle for \dzdstm channel is also studied.
The helicity angle, $\Theta$, is defined as the angle between
the direction opposite to the $B$ meson and that of the slow pion in the 
the $D^{*-}$ rest frame. Figure \ref{fig:helang} shows
the $\de$ sideband subtracted (the signal region is defined as $|\de|<$20 MeV 
and sideband regions are 50 MeV $< |\de| <$ 70 MeV) helicity angle 
distributions in data and for the MC signal. The data
follow a $\cos^2 \Theta$ distribution as expected for a 
$B$ to pseudoscalar-vector decay.

The systematic uncertainty, shown in Table \ref{tab:system}, is obtained 
from a quadratic sum of	the uncertainties in 
(a) the track finding efficiency, ranging from 1\% 
    for high momentum tracks to 8\% 
    for pions of 80 MeV/$c$, estimated from partially reconstructed 
    $D^{*-} \to \overline{D}{}^0 (\to K_S (\to \pi^+ \pi^-) \pi^+ \pi^-) \pi^-$
    events and a track embedding study; 
(b) the $\pi^0$ reconstruction efficiency, estimated from a comparison of 
    $D^0 \to K^- \pi^+ \pi^0$ yields in data and MC ; 
(c) the $K_S$ selection efficiency, estimated from a comparison of
    $D^0 \to K_S \pi^+ \pi^-$ yields in data and MC; 
(d) $K/\pi$ selection efficiencies, estimated using 
    $D^{*-} \to \overline{D}{}^0 (\to K^+ \pi^-) \pi^-$ events; 
(e) The world-average $D^0, D^+$ and $D^{*+}$ branching fractions ~\cite{pdg2004};
(f) scaling factor for $\de$ distributions in data; 
(g) MC statistics; 
(h) the total number of ${B\overline{B}}$ events ($N_{B\overline{B}}$). 
The systematic error is also studied by 
(a) varying the \de fit ranges within $-$200 MeV to +200 MeV; 
(b) the choice of the fitting functions;
(c) deriving the branching fraction without selection of the best candidate. 
Systematic uncertainties from the latter three sources are  
negligible compared to those from the other sources.

\begin{table}[htbp] 
 \begin{center}
 \caption{Individual contributions of systematic uncertainties (in \%).}
 \begin{ruledtabular}
  \begin{tabular}{lccc}
                        &  $D^+ D^-$ & $D^0 D^-$ & $D^0 D^{*-}$ \\ \hline
Track finding           & 6.5 &  7.2      & 9.1 \\
$\pi^0$                 & -   & -         & 2.0 \\
$K_S$                   & -   & 0.4       & 1.0 \\
K/$\pi$ selection       & 5.5 & 5.5       & 5.1 \\
$\de$ scale             & 2.4 & 2.6       & 2.2 \\ 
MC statistics           & 0.9 &1.2        & 2.0 \\
$N_{B\overline{B}}$     & 0.5 &0.5        & 0.5 \\ 
$D$ branching fractions     & 12.1&6.6        & 5.3 \\ \hline
Total                   & 15.0& 11.5      &12.1  \\
\end{tabular}
\end{ruledtabular}
\label{tab:system}
\end{center}
\end{table}

Charge asymmetry, $A = ({N_- - N_+})/({N_- + N_+})$ in 
$\dzerdm $ and $\dzdstm $ channels is $-$0.05 $\pm$ 0.15 $\pm$ 0.05
and 0.15 $\pm$ 0.15 $\pm$ 0.05, respectively, where $N_-(N_+)$ is the 
number of observed events in $B^- (B^+)$ decays. 
Systematic errors on charge asymmetries are determined from
high statistics $B^- \to D^{(*)0}D_s^{(*)-}$ and $B \to D^* (n\pi)$ event
samples.

In summary, we report first observations of the decays  
\dpludm, \dzerdm and \dzdstm using 152 million 
$ B \overline{B}$ events. We measure branching fractions for 
these three decay modes, which are consistent with the expectation 
from SU(3) symmetry. Charge asymmetries in the $\dzerdm $ and 
$\dzdstm $ are consistent with zero. 
We find a statistically significant signal in the $\dpludm$ channel;
in the future this mode will be used to perform additional studies of 
time dependent \CP violation in $b \to c \overline{c} d$ decays.

We thank the KEKB group for the excellent operation of the
accelerator, the KEK cryogenics group for the efficient
operation of the solenoid, and the KEK computer group and
the NII for valuable computing and Super-SINET network
support.  We acknowledge support from MEXT and JSPS (Japan);
ARC and DEST (Australia); NSFC (contract No.~10175071,
China); DST (India); the BK21 program of MOEHRD and the CHEP
SRC program of KOSEF (Korea); KBN (contract No.~2P03B 01324,
Poland); MIST (Russia); MESS (Slovenia); Swiss NSF; NSC and MOE
(Taiwan); and DOE (USA).


\begin{thebibliography}{99}

\bibitem{cpsm} Belle Collaboration, K. Abe {\it et al.,} 
        Phys. Rev. D{\bf 66}, 071102(R) (2002), 
%        BaBar Collaboration, B. Aubert 
%        {\it et al.,}  Phys. Rev. Lett. {\bf  90}, 221801 (2003), 
	BaBar Collaboration, B. Aubert  
        {\it et al.,} Phys. Rev. Lett. {\bf  89}, 201802 (2002). 
\bibitem{beyond} Y. Grossman and M. Worah, Phys. Lett. B{\bf 395}, 
        241 (1997).
\bibitem{sanda} A.I. Sanda and Zhi-zhong Xing, Phys. Rev. D{\bf 56},
        341 (1997).
\bibitem{cpvxx} Belle Collaboration, K. Abe {\it et al.,} 
        Phys. Rev. Lett. {\bf 89}, 122001 (2002), 
        BaBar Collaboration, B. Aubert {\it et al.,} 
        Phys. Rev. Lett. {\bf 89}, 061801 (2002), BaBar Collaboration, 
        B. Aubert {\it et al.,} Phys. Rev. Lett. {\bf 90}, 221801 (2003),
	Belle Collaboration, T. Aushev {\it et al.,} Phys. Rev. Lett. {\bf 93}, 
        201802 (2004),
        Belle Collaboration, H. Miyake {\it et al.,} Submitted to Phys. Lett. {\bf B}.
\bibitem{pdg2004} S. Eidelman {\it et al.,} (Particle Data Group), 
                   Phys. Lett. B{\bf 592}, 1 (2004).

\bibitem{belledet} Belle Collaboration, A. Abashian {\it et al.,} 
                   Nucl. Instr. Meth. A{\bf 479}, 117, (2002).
\bibitem{kekb}     S. Kurokawa and E. Kikutani, 
                   Nucl. Instr. Meth. A{\bf 499}, 1 (2003).
\bibitem{foxwol}   G. C. Fox and S. Wolfram, 
                   Phys. Rev. Lett. {\bf 41} (1978) 1581.
\bibitem{qq98}     The $QQ$ $B$ meson decay event generator was developed by 
                   the CLEO Collaboration, 
                   http://www.lns.cornell.edu/public/CLEO/soft/QQ
\bibitem{geant3}   CERN Program Library Long Writeup, W5013, CERN, 1993.

\end{thebibliography}
\end{document}